\documentclass[aps,pra,superscriptaddress,twocolumn,showpacs,noeprint]{revtex4-1}
\usepackage{graphicx,amsfonts,times,bm,amsmath,verbatim,color,array}

\usepackage{natbib}
\usepackage{graphicx,amssymb,amsmath,ifthen,braket,xcolor}
\usepackage{bm}

\begin{document}

\title{Flux-driven quantum spin liquids in kagome optical lattices}

\author{Hoi-Yin Hui}
\thanks{Current address: Blueshift Asset Management, 151 Bodman Pl., Suite 301, Red Bank, New Jersey 07701, United States}
\affiliation{Department of Physics, Virginia Tech, Blacksburg, Virginia 24061, USA}

\author{Mengsu Chen}
\affiliation{Department of Physics, Virginia Tech, Blacksburg, Virginia 24061, USA}

\author{Sumanta Tewari}
\affiliation{Department of Physics and Astronomy, Clemson University, Clemson, South Carolina 29634, USA}

\author{V.W. Scarola}
\email[Email address:]{scarola@vt.edu}
\affiliation{Department of Physics, Virginia Tech, Blacksburg, Virginia 24061, USA}

\begin{abstract}
Quantum spin liquids (QSLs) define an exotic class of quantum ground states where spins are disordered down to zero temperature. We propose routes to QSLs in kagome optical lattices using applied flux.  An optical flux lattice can be applied to induce a uniform flux and chiral three-spin interactions that drive the formation of a gapped chiral spin liquid.  A different approach based on recent experiments using laser assisted tunneling and lattice tilt implements a staggered flux pattern which can drive a gapless spin liquid with symmetry protected nodal lines.  Our proposals therefore establish kagome optical lattices with effective flux as a powerful platform for exploration of QSLs.
\end{abstract}


\maketitle

\section{Introduction}
QSLs are highly entangled spin states that are quantum disordered down to zero temperature and therefore do not display conventional features of magnetism.  But QSLs may nonetheless offer explanations for strongly correlated phenomena observed in some materials \cite{Balents2010,Zhou2017}.  Frustration is known to favor certain types of QSLs. Kagome lattice models of spins in particular serve as a central archetype hosting a broad array of QSLs.  It is now well established that ground states arising from the standard antiferromagnetic Heisenberg interaction ($\bm S_i \cdot \bm S_j$, where $\bm S_i$ is the usual spin operator at a site $i$) on  a kagome lattice can be driven into exotic spin liquids when certain three-spin interactions [$\bm S_i\cdot(\bm S_j\times\bm S_k)$] are added to the Heisenberg interaction \cite{Kalmeyer1987a,Wen1989,Wen1990,Greiter2014,Bauer2014,Kumar2015,Bauer2018a}.

When the three-spin interaction is added uniformly everywhere to the kagome lattice, a chiral spin liquid (CSL) arises \cite{Wen1989,Wen1990,Bauer2014} since it is an exact ground state of similar interactions \cite{Schroeter2007,Greiter2014}.  A CSL is related to a bosonic Laughlin state \cite{Laughlin1983b,Kalmeyer1987a,Greiter2014}, and, as such, derives some of the same properties.  The CSL is a topologically ordered ground state and is therefore two-fold degenerate on the torus.   Such a topological degeneracy is a key feature of gapped topologically ordered states that can be used to uniquely identify them in numerics \cite{Haldane1985c,Haldane1985e,Wen1989b,Wen1990a}.  The CSL also possesses chiral edge modes; It derives from flux attachment in effective Chern-Simons theories \cite{Kumar2015}; And furthermore, the CSL hosts exotic anyon excitations, whereby braiding of anyons changes the many-body wavefunction by a non-trivial phase \cite{Wilczek1982,Nayak2008}.  Identifying such exotic braid statistics in the laboratory is a key goal of quantum many-body physics \cite{Nayak2008}.

\begin{figure}
	\begin{center}
		\includegraphics[width=0.48\textwidth]{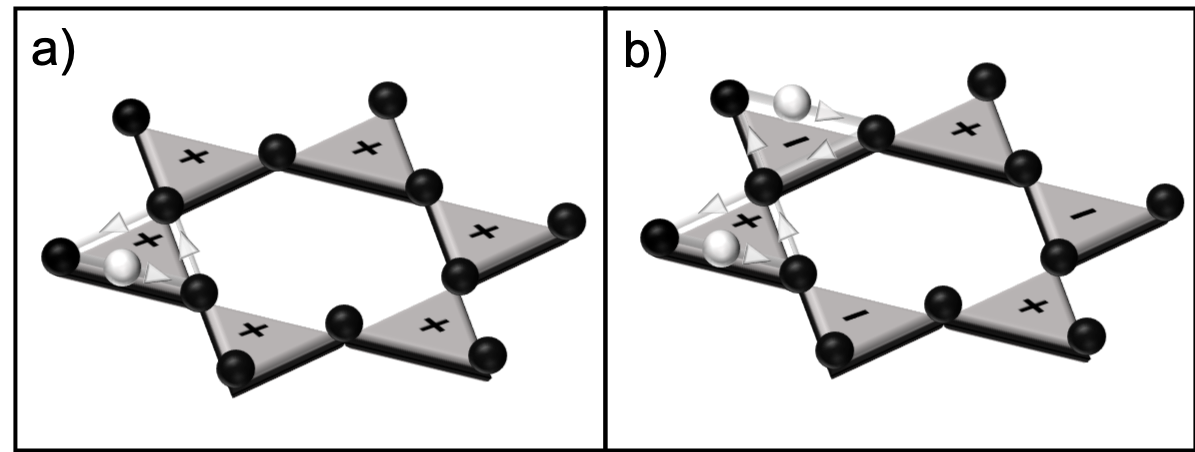}
	\end{center}
	\caption{Schematic of one fermion per site (blacks spheres) on a kagome lattice.  The white spheres denote virtual current driven by flux passing through the lattice.  The plus and minus signs denote the sign of the flux captured by the virtual currents.  Panel a (b) shows a uniform (staggered) flux pattern.  In the Heisenberg limit of a Hubbard model, virtual currents encircling flux lead to chiral three-spin terms that drive spin liquids.  }\label{fig_schematics_kagome_flux}
\end{figure}

Prospects for driving kagome antiferromagnets into the CSL remains daunting and rare in the published literature.  Recent works with ultracold atoms placed in optical lattices \cite{Bloch2008,Gross2017} show promise because not only are kagome lattices possible \cite{Santos2004,Jo2012a}, but also temperatures low enough to realize antiferromagnetic order derived from super exchange between fermionic atoms have recently been realized \cite{Mazurenko2017} with atomic gas microscopes \cite{Bakr2009,Sherson2010,Bakr2010,Endres2011,Weitenberg2011a,Islam2015,Hild2014,Preiss2015,Cheuk2015,Parsons2015,Haller2015,Miranda2015a,Yamamoto2016,Parsons2016,Boll2016,Cheuk2016,Choi2016,Drewes2017}.  One recent idea suggests that a CSL may be realizable in systems of polar molecules using long-ranged dipolar interactions in optical lattices \cite{Yao2018}.  In this paper we examine a very different approach based on more common short-ranged interactions of fermionic atoms in the presence of tunable fluxes. 

We model fermionic atoms placed in kagome optical lattices with flux.  Effective flux in optical lattices can be realized in a variety of ways \cite{Williams2010a,Aidelsburger2011,Cooper2011,Cooper2011b,Jimenez-Garcia2012,Struck2012,Juzeliunas2012a,Hauke2012,Juzeliunas2012a,Struck2013,Aidelsburger2013,Miyake2013,Cooper2013,Jotzu2014a}. We show that an optical flux lattice \cite{Cooper2011,Cooper2011b,Juzeliunas2012a,Cooper2013} can be used to generate a sufficient amount of flux to perturbatively drive virtual currents in an underlying Hubbard model \cite{Rokhsar1990,Sen1995,Scarola2004}.   Fig.~\ref{fig_schematics_kagome_flux}a shows one fermion per site in the Mott limit.  Ordinary hopping is prevented but virtual hops around triangles can capture flux to drive three-spin terms needed to enhance the CSL. The equivalent amount of flux for such terms in a solid with an $\sim 1$ \AA~  inter-atomic spacing would require large magnetic fields, $\sim 10^4$ T.  We will therefore show that an optical flux lattice in a kagome optical lattice offers a more direct route to the CSL than what is achievable in solids with ordinary magnetic field strengths. 

We also model effective flux generated by laser assisted tunneling combined with a potential tilt as first implemented in square optical lattices \cite{Aidelsburger2013,Miyake2013}.  When examining this setup in a kagome optical lattice we find that the effective flux pattern is staggered (Fig.~\ref{fig_schematics_kagome_flux}b).  We speculate that this flux pattern may be able to drive an interesting gapless spin liquid recently discovered numerically \cite{Bauer2018a} to host symmetry protected nodal lines and may thus offer a platform to study gapless spinon surfaces \cite{Motrunich2007,Sheng2009,Block2011,Jiang2013a}.  Overall, we show that flux applied to kagome optical lattices offers a powerful tool to study QSLs, in particular, the long sought chiral spin liquid, and, possibly, a spin liquid with gapless spinon surfaces.   

This paper is organized as follows. In Sec.~\ref{sec:hubb} we discuss the well-known Hubbard limit of kagome optical lattices.  We generalize to the case of complex hopping and describe a calculation showing that, for large optical lattice depths (weak hopping), the interactions lead to spin models with interesting three-body terms that drive spin liquid formation.  This section describes a calculation we will use in the remainder of the paper. In Sec.~\ref{sec:uniform} we discuss an optical flux lattice setup that leads to a uniform effective flux (Fig.~\ref{fig_schematics_kagome_flux}a).  We use the derivation in Sec.~\ref{sec:hubb} to argue that the optical flux lattice establishes three-spin terms favoring a CSL.  In Sec.~\ref{sec:staggered} we discuss a route to introduce a staggered flux (Fig.~\ref{fig_schematics_kagome_flux}b) with laser assisted tunneling and a lattice tilt.  We again use Sec.~\ref{sec:hubb} to argue for a spin model with three-spin terms.  Although here we find that the resulting three-spin terms are staggered and relate to recent work on gapless spin liquids with nodal spinon surfaces \cite{Bauer2018a}.  We end with a summary and conclusion in Sec.~\ref{sec:discussion}, where we also discuss discuss practical aspects: entropy requirements and methods to observe these spin liquid phases.

\section{Hubbard model and kagome lattices}\label{sec:hubb}

We begin by discussing the mathematical connection between effective spin models and Hubbard models in the presence of flux. Secs.~\ref{sec:uniform} and ~\ref{sec:staggered}  will rely on the derivation here as a route to model two distinct proposals to realize effective flux in kagome optical lattices.  In both cases we assume fermionic alkali atoms equally populating the two lowest  hyperfine states to yield a pseudospin.  We also assume that they are loaded into a kagome optical lattice \cite{Santos2004,Ruostekoski2009,Jo2012a}.  The details of the kagome optical lattice setup have been discussed elsewhere \cite{Jo2012a}, where it was found that overlaying two triangular optical lattices formed from lasers with commensurate wavelengths yield potentials deep enough to realize the Hubbard limit \cite{Jaksch1998,Santos2004,Jo2012a}.  For laser intensities yielding a Bloch bandwidth well below the band gap, we have \cite{Jaksch1998,Santos2004,Jo2012a}: 
\begin{equation}
H^{\alpha}=H_0^{\alpha}+U\sum_i n_{i\uparrow}n_{i\downarrow},
\label{eq_Htotal}
 \end{equation}
 where the second term is a repulsive Hubbard interaction originating from the $s$-wave scattering between atoms in spin states $\sigma\in \{\uparrow,\downarrow\}$.  Here $n_{i\sigma}=
 a_{i\sigma}^\dag a_{i\sigma}^{\vphantom{\dag}}$
 is defined in terms of dressed fermion annihilation ($a_{j\sigma}^{\vphantom{\dag}}$)  and creation ($a_{i\sigma}^\dag$) operators at the site $\bm R_i$.  The first term is a single particle hopping term:
  \begin{equation}
 H_0^{\alpha}=-\sum_{\left<ij\right>}t_{ij}^{\alpha}a_{i\sigma}^\dag a_{j\sigma}^{\vphantom{\dag}},
 \label{eq_H_0_end}
  \end{equation}
 with nearest neighbor hopping matrix elements $t_{ij}^{\alpha}$.   Eq.~\ref{eq_Htotal} defines the essential degrees of freedom we will examine. 
 
  We will discuss two different strategies to realize effective magnetic fields strong enough to drive Mott insulating states toward QSLs in kagome optical lattices.   The first strategy, discussed in Sec.~\ref{sec:uniform}, will examine the optical flux lattice as a route to a uniform flux pattern,  Fig.~\ref{fig_schematics_kagome_flux}a, $\alpha=\text{Un}$.  The second strategy, discussed in Sec.~\ref{sec:staggered}, will examine laser assisted tunneling combined with a potential tilt, as a route to implement a staggered flux lattice, Fig.~\ref{fig_schematics_kagome_flux}b, $\alpha=\text{St}$.  In both cases the flux can be described by effective gauge fields, $\bm A$, captured by a complex hopping via the Peierls  transformation: $t_{ij}^{\alpha}= \vert t_{ij}\vert \exp(i\Phi_{ij}/\Phi_0)$, where the flux on a bond is $\Phi_{ij}=\int_{\bm R_i}^{\bm R_j} \bm A\cdot d\bm r$.  The flux then leads to an Aharonov-Bohm phase difference as a particle tunnels around a triangle: $2\pi\Phi_{\Delta}/\phi_0$, where $\Phi_{\Delta}=\int_{\Delta}(\bm\nabla\times\bm A)\cdot d^2\bm r$ is the flux through an upward pointing triangle in the kagome lattice ($\Phi_{\nabla}$ is defined in the same way but for downward pointing triangles).  In the following we work in units $\hbar=a=q=1$ where $a$ is the lattice spacing and $q$ is the effective charge, so that $\Phi_0=2\pi$.   
  
We now turn to interaction effects in the Heisenberg limit to study the role of our proposed flux patterns in driving QSLs.  Eq.~\ref{eq_Htotal} is well approximated by spin models when there is one particle per site and for $t\ll U$.  In this limit we can derive the spin model by expanding  $H^{\alpha}$ perturbatively in powers of $t/U$ using $\exp(iK) H^{\alpha} \exp(-iK)$ where $K$ is an operator that changes the number of doubly occupied sites \cite{MacDonald1988}.  Projecting into the limit of one particle per site we have \cite{MacDonald1988,Rokhsar1990,Sen1995}:
\begin{eqnarray}
 &H^{\alpha}&\approx J_{\text{H}}\sum_{\left<ij\right>} \bm S_i\cdot \bm S_j 
+J_{\text{C}} (\Phi) \bigl [
\sum_{ijk\in\bigtriangleup}\bm S_i\cdot(\bm S_j\times\bm S_k)
\nonumber \\
&+&P_{\alpha}\sum_{ijk\in\bigtriangledown}\bm S_i\cdot(\bm S_j\times\bm S_k)
\bigr] + \mathcal{O}(t^4/U^3)
\label{eq_spin_model}
\end{eqnarray}
where we have used the mapping:
 $\bm S_i =(1/2) a_{i\sigma}^\dag \bm \sigma_{\sigma,\sigma'} a_{i\sigma'}^{\vphantom{\dag}}$, with $\bm \sigma_{\sigma,\sigma'}$ the elements of the usual Pauli matrices.  The first term is the usual antiferromagnetic Heisenberg term arising from 2 virtual hops along bonds: $J_{\text{H}}=4t^2/U$.  Here we assumed that the magnitude of the hopping on all bonds, $t$, is the same without loss of generality.

In the absence of flux, corrections to the usual Heisenberg (two-spin) interaction are fourth-order (four-spin) terms.  But here we note that three-spin terms in Eq.~\ref{eq_spin_model} arise perturbatively from third-order virtual hops around triangles due to the presence of an effective field.  One can see that they are non-zero only in the presence of time-reversal symmetry breaking on individual triangles due to effective fluxes: $J_{\text{C}}(\Phi)=(24 t^3/U^2)\vert \sin(2\pi\Phi_{\Delta,\nabla}/\Phi_0)\vert $, since $J_{\text{C}}(\Phi)$ vanishes at zero flux. In the following we seek routes to impose the maximum amount of flux through each triangle:
$\vert \Phi_{\Delta,\nabla}\vert=\Phi_0/4$ to maximize the strength of the three-spin terms.  The parameter $P_\alpha$ captures both the uniform flux  $P_{\alpha=\text{Un}}=1$ ($\Phi_{\Delta}=\Phi_{\nabla}$) and the staggered flux cases $P_{\alpha=\text{St}}=-1$ ($\Phi_{\Delta}=-\Phi_{\nabla}$) discussed below.  

 The lowest order corrections to Eq.~\ref{eq_spin_model} arise from fourth order virtual hops.  Corrections of order $t^4/U^3$ modify $J_{\text{H}}$.  In addition to modifying $J_{\text{H}}$, fourth order corrections also give rise to next-nearest neighbor Heisenberg terms, e.g., $\bm S_i\cdot \bm S_{i+2}$.  We exclude next-nearest neighbor Heisenberg terms in the following.  

The following two sections propose to realize both uniform and staggered fluxes, and therefore Eqs.~\ref{eq_Htotal}~-~\ref{eq_spin_model}, in kagome optical lattices.  We will discuss how each separate setup can be approximated by Eq.~\ref{eq_H_0_end}, but with different flux patterns.  We discuss the role of interactions by using the derivation of the above general spin model to argue that a Mott insulator placed in a deep optical lattice is approximated by special cases of Eq.~\ref{eq_spin_model}.   The uniform flux case ($\alpha=\text{Un}$) will yield a three-spin term in Eq.~\ref{eq_spin_model} that is the same on all triangles in the lattice (Sec.~\ref{sec:uniform}).  The staggered flux case ($\alpha=\text{St}$) will yield a three-spin term in Eq.~\ref{eq_spin_model} with a sign that alternates from triangle to triangle (Sec.~\ref{sec:staggered}). 

\section{Uniform Flux from an optical flux lattice}\label{sec:uniform}

Optical flux lattices \cite{Cooper2011} offer a straightforward route to implement a uniform effective flux in a kagome optical lattice.  In optical flux lattices proposed so far, the lowest (of two) hyperfine states adiabatically evolves under external lasers so that the Berry's phase of an atom adiabatically changes around closed loops in a lattice to mimic an effective magnetic field. Details of different optical flux lattices have been discussed in the literature \cite{Cooper2011,Cooper2011b,Juzeliunas2012a,Cooper2013}.  A particularly versatile setup was proposed in Ref.~\cite{Cooper2011b} wherein a two-photon dressed state can be used to address many common atomic species.  There it was argued that low-loss fermionic atoms which have already been laser cooled, such as $^{171}\text{Yb}$ can $^{199}\text{Hg}$, can be used.  By loading them into the lowest two hyperfine levels and addressing with two lasers detuned from the first excited level one can effect a Berry's phase change.  The low energy states in this proposal \cite{Cooper2011b} yield just a  single-component fermion moving in an effective magnetic field.

To generate effective flux for a system of two-component fermions (a spin model), we consider a straightforward two-copy generalization of Ref.~\cite{Cooper2011b} where the atoms are loaded into four near-degenerate lowest levels (as opposed to just two).  Beams implementing the optical flux lattice are similarly detuned from excited states.   The resulting four hyperfine states reduce to a dressed state of just the two lowest levels thus leading to an effective spin in an optically induced field.  The dynamics of each atom leads to a Berry's phase \cite{Cooper2011} which is equivalent to a flux passing through a closed loop, identical for each of the two lowest hyperfine states.  Candidate atoms include isotopes of alkaline-earth-like atoms, such as $^{173}\text{Yb}$ and $^{87}\text{Sr}$ \cite{Fukuhara2007,DeSalvo2010,Tey2010} which can be used to prepare $SU(N)$-symmetric Hubbard models.  Recent work has been able to use optical pumping to load $^{137}\text{Yb}$ into four degenerate lowest levels and cool into a Mott insulator displaying spin correlations \cite{Ozawa2018a}. These four levels can be split with a Zeeman field to yield an excellent candidate for a two-copy generalization of Ref.~\cite{Cooper2011b}. 

\begin{figure}
	\begin{center}
		\includegraphics[width=0.48\textwidth]{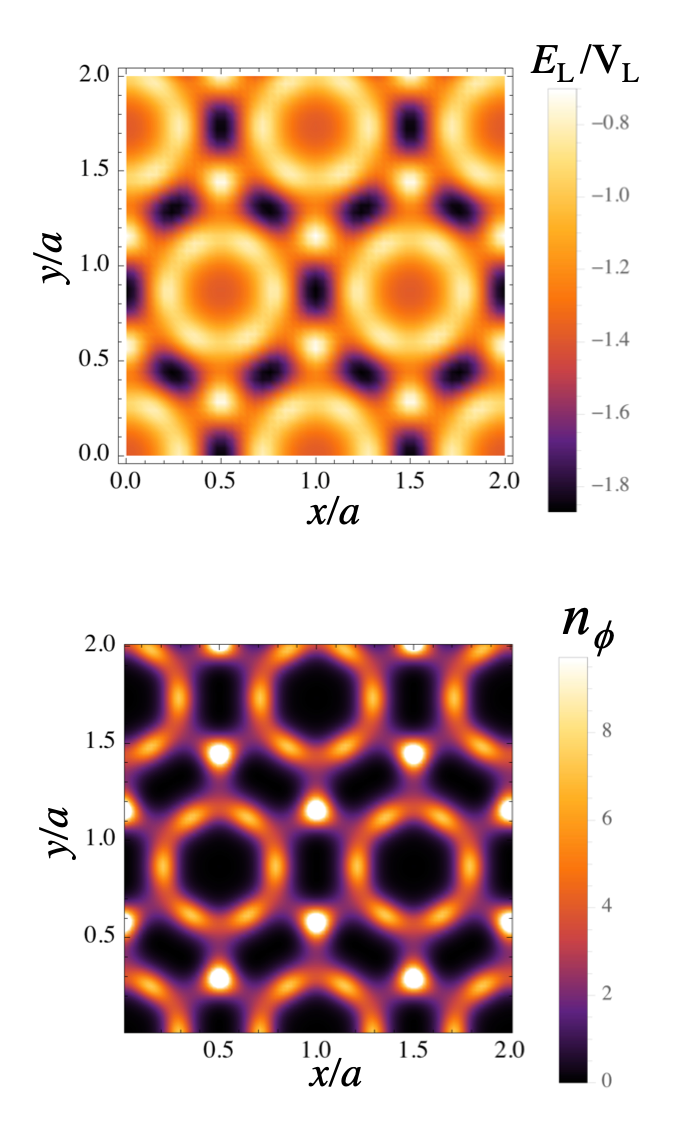}
	\end{center}
	\caption{Top: Lowest energy of the potential term in Eq.~\ref{eq:ofl}, $E_{\text{L}}$, used to implement the optical flux lattice for lattice depth $V_L=1$.  Position is plotted in units of the lattice spacing $a$. The atoms sit at the energy minima (dark regions).  Bottom: Same as the top but for the magnitude of the effective flux density, $n_\phi$. The bright spots within triangles show that atoms experience a uniform flux through triangles (Fig.~\ref{fig_schematics_kagome_flux}a). }\label{fig_flux_lattice}
\end{figure}

We first examine the non-interacting part of the kagome optical flux lattice. The optical flux lattice arises from counter propagating lasers defining the usual kagome potential but for the four hyperfine states so that, for each pair of hyperfine states, the single-particle Hamiltonian becomes:
\begin{equation}
\tilde{H}_0^{\text{Un}}= \frac{\bm p^2}{2m}+
V_L\sum_{l=1}^3\left[\cos\left(\bm k_l\cdot\bm r\right)-\frac{1}{5}\cos\left(2\bm k_l\cdot\bm r\right)\right]\sigma_l,
\label{eq:ofl}
\end{equation}
where $V_L$ is the lattice depth, $\bm k_1=(0,1)$, $\bm k_2=\left(\sqrt{3}/2,-1/2\right)$, $\bm k_3=\left(-\sqrt{3}/2,-1/2\right)$, and $\sigma_l$ are the Pauli matrices. If the kinetic energy is much smaller than the gap of the second term, the ground state adiabatically follows the second term in a dressed state $\phi(\bm r)$.  Writing the ground eigenstate of the second term as  $\vert \Psi\rangle^{\dagger}=(\phi_1(\bm r),\phi_2(\bm r))$ we 
assume that the ground state of this Hamiltonian is non-degenerate everywhere.  Projecting to its lower band leads to an effective two-component Hamiltonian with the vector potential
$
\bm A=i\left<\Psi\right|\nabla_{\bm r}\left|\Psi\right>,
$
where the effective magnetic flux density perpendicular to the plane of the lattice becomes
$n_{\phi}\equiv(\bm\nabla\times\bm A)\cdot \hat{z}/\Phi_0$.  

The top panel of Fig.~\ref{fig_flux_lattice} plots the lowest energy of the potential term in Eq.~\ref{eq:ofl}, $E_L$, for a single spin.  Here we see that minima correspond to a kagome lattice as expected.  The bottom panel plots the flux density in the lattice.  The flux density pattern shows that the flux piercing each triangle is the same, thus corresponding to Fig.~\ref{fig_schematics_kagome_flux}a.

The flux through the lattice can be tuned to yield a complex hopping.  Passing to the tight binding limit we assume that $V_L$ is large enough to keep all atoms in the lowest band of the kagome lattice (more than a few atomic recoils).  Eq.~\ref{eq:ofl} then becomes well approximated by Eq.~\ref{eq_H_0_end} with complex hopping, $t_{ij}=t \exp(i\Phi_{ij}/\Phi_0)$, where $t$ is real and the same for all bonds.  The hoppings capture a uniform flux passing through all triangles in the kagome lattice ($\Phi_{\Delta}=\Phi_{\nabla}$).  We have checked that the flux passing through triangles in  Fig.~\ref{fig_flux_lattice} is maximized: 
$\text{Im}[ t_{12} t_{23} t_{31} ]= t^3 \sin(2\pi\Phi_{\Delta}/\Phi_0)= t^3 \sin(2\pi\Phi_{\nabla}/\Phi_0)\approx t^3$. This shows that an optical flux lattice can be tuned to yield a large uniform effective flux through a kagome optical lattice captured by the Hubbard model discussed in Sec.~\ref{sec:hubb}.

\begin{figure}
	\begin{center}
		\includegraphics[width=0.48\textwidth]{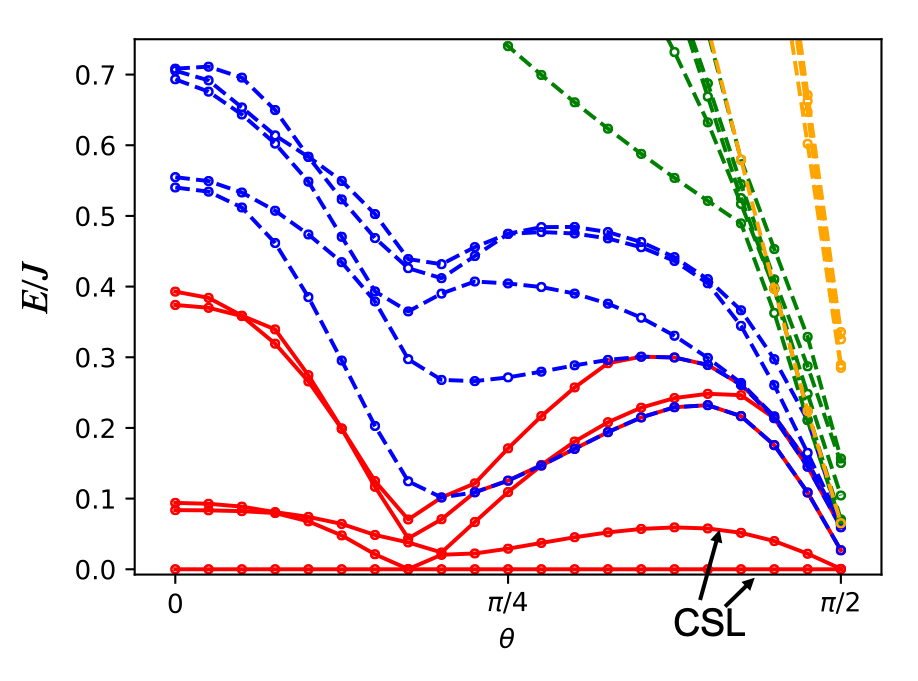}
	\end{center}
	\caption{Lowest energy eigenvalues of the spin model with uniform effective flux (Eq.~\ref{eq_spin_model_uniform}) for 18 spins in $2\times3$ unit cells with periodic boundaries. The lowest energy is set to zero.  The strengths of the Heisenberg and chiral three-spin term are parameterized with $J_{\text{H}}=J\cos(\theta)$ and $J_{\text{C}}(\Phi)=J\sin(\theta)$, respectively.  With this parameterization, we have $\tan(\theta)=6t/U$ and $\Phi_{\Delta}=\Phi_{\nabla}=\Phi_0/4$ in the original Hubbard model.  The right side of the graph is dominated by the three-spin term where we see two ground states defining the CSL.  Different colors indicate different total spin sectors: $S_z = 0$ (red), $S_z = 1$ (blue), $S_z = 2$ (green), and $S_z = 3$ (orange).  }\label{fig_energy_uniform_flux}
\end{figure}

 We now turn to interaction effects in the uniform flux case.  We assume the Hubbard limit with one particle per site.   Perturbation theory discussed in Sec.~\ref{sec:hubb} with complex hopping leads to a spin model of the form:
 \begin{eqnarray}
 &H^{\text{Un}}&\approx J_{\text{H}}\sum_{\left<ij\right>} \bm S_i\cdot \bm S_j 
+J_{\text{C}}(\Phi_0/4)
\sum_{ijk\in\{\bigtriangleup,\bigtriangledown\}}\bm S_i\cdot(\bm S_j\times\bm S_k) \nonumber \\
 &+& \mathcal{O}(t^4/U^3)
\label{eq_spin_model_uniform}
\end{eqnarray}
 where we assumed a uniform flux, $\Phi_{\Delta}=\Phi_{\nabla}=\Phi_0/4$, that leads to the three-spin interaction term that is uniform throughout the lattice and $J_{\text{C}}(\Phi_0/4)=24t^3/U^2$.

We expect Eq.~\ref{eq_spin_model_uniform} to lead to a gapped CSL \cite{Kalmeyer1987a,Schroeter2007,Greiter2014,Bauer2014}.  The three-spin term on the kagome lattice, and therefore large $J_{\text{C}}(\Phi)\sim t^3/U^2$ in Eq.~\ref{eq_spin_model_uniform}, strongly favors the CSL.  But the derivation of Eq.~\ref{eq_spin_model_uniform} is most accurate in the perturbative limit, $t/U \ll 1$.  We therefore search for an intermediate range of $t/U$ which lies in the perturbative regime while still favoring the CSL.

We study the robustness of the CSL over the entire parameter range using unbiased exact diagonalization for small system sizes.  We use the Krylov-Schur algorithm \cite{Stewart2001} which allows us to handle degenerate eigenvalues.  This method is essentially exact (it includes all quantum fluctuations) and gives the same results as other unbiased methods on small lattices.  We work on a finite system size, 18 spins ($2\times3$ unit cells) with periodic boundaries to obtain the lowest energy states.  We point out that similar small-system size studies in the fractional quantum Hall regime on related models \cite{Jain2007} are applicable to the thermodynamic limit because correlations in gapped topological phases are known to decay exponentially.  For example, system sizes as small as 8 particles capture the low-energy roton structure of the fractional quantum Hall states \cite{Haldane1985c}.  Since the CSL maps to the bosonic fractional quantum Hall states, the presence of the gap and other corroborating numerics \cite{Bauer2014} allow us to make conclusions about the robustness of the CSL.

 Figure~\ref{fig_energy_uniform_flux} plots the lowest energies of Eq.~\ref{eq_spin_model_uniform} as a function of the relative strength of each term using exact diagonalization.  To make a more compact exploration of parameter space, we define new interaction parameters, $J$ and $\theta$, via $J_{\text{H}}=J\cos(\theta)$ and $J_{\text{C}}(\Phi_0/4)=J\sin(\theta)$, corresponding to $\tan(\theta)=6t/U$ and $\Phi_{\Delta}=\Phi_{\nabla}=\Phi_0/4$ in the original Hubbard model.   By varying $\theta$ in Fig.~\ref{fig_energy_uniform_flux} we can tune between the Heisenberg (left) and three-spin (right) limits in Eq.~\ref{eq_spin_model_uniform}.  The rightmost side of the graph shows a two-fold degenerate ground state (arrows), as expected for a CSL on a torus.  There is a gap to a third state that remains robust for $\theta \gtrsim \pi/4$, i.e., $J_{\text{C}}/J_{\text{H}}\gtrsim 0.5$.  In this regime we see that deviations from the exact CSL generating model \cite{Greiter2014} lift the exact degeneracy induced by the three-spin term at $\theta=\pi/2$.  Nonetheless the CSL remains robust since the gap does not close.  The gap is even somewhat enhanced by the two-spin term.  Larger system size numerics \cite{Bauer2014} show an even larger range of stability,  $J_{\text{C}}/J_{\text{H}}\gtrsim 0.15$ in the thermodynamic limit.  
 
 Returning to the original Hubbard parameters, this range of CSL stability, $t/U\gtrsim 0.16$ for 18 spins and $t/U \gtrsim 0.026$ in the thermodynamic limit, corresponds to parameters well within the assumption of the perturbative regime.  This brings us to our central result: an optical flux lattice induces third-order virtual currents which in turn drive a CSL state in the Mott insulator regime of a Hubbard-kagome optical lattice.   We discuss possible routes to observation of the CSL in Sec.~\ref{sec:discussion}.  The next section discusses a method to introduce staggered flux and possibly a different QSL in a kagome optical lattice.

\section{Staggered Flux from a Moving Optical Lattice}\label{sec:staggered}
  
We now discuss a separate method to introduce flux in a kagome optical lattice.  The method is based on a scheme recently used to implement complex hopping in a square optical lattice \cite{Miyake2013,Aidelsburger2013} with a tilt (which can be applied using a variety of methods including a magnetic field or gravity) and a moving lattice.  The moving lattice is established by two additional Raman lasers applied perpendicular to the tilt. 

\begin{figure}[t]
	\begin{center}
		\includegraphics[width=0.48\textwidth]{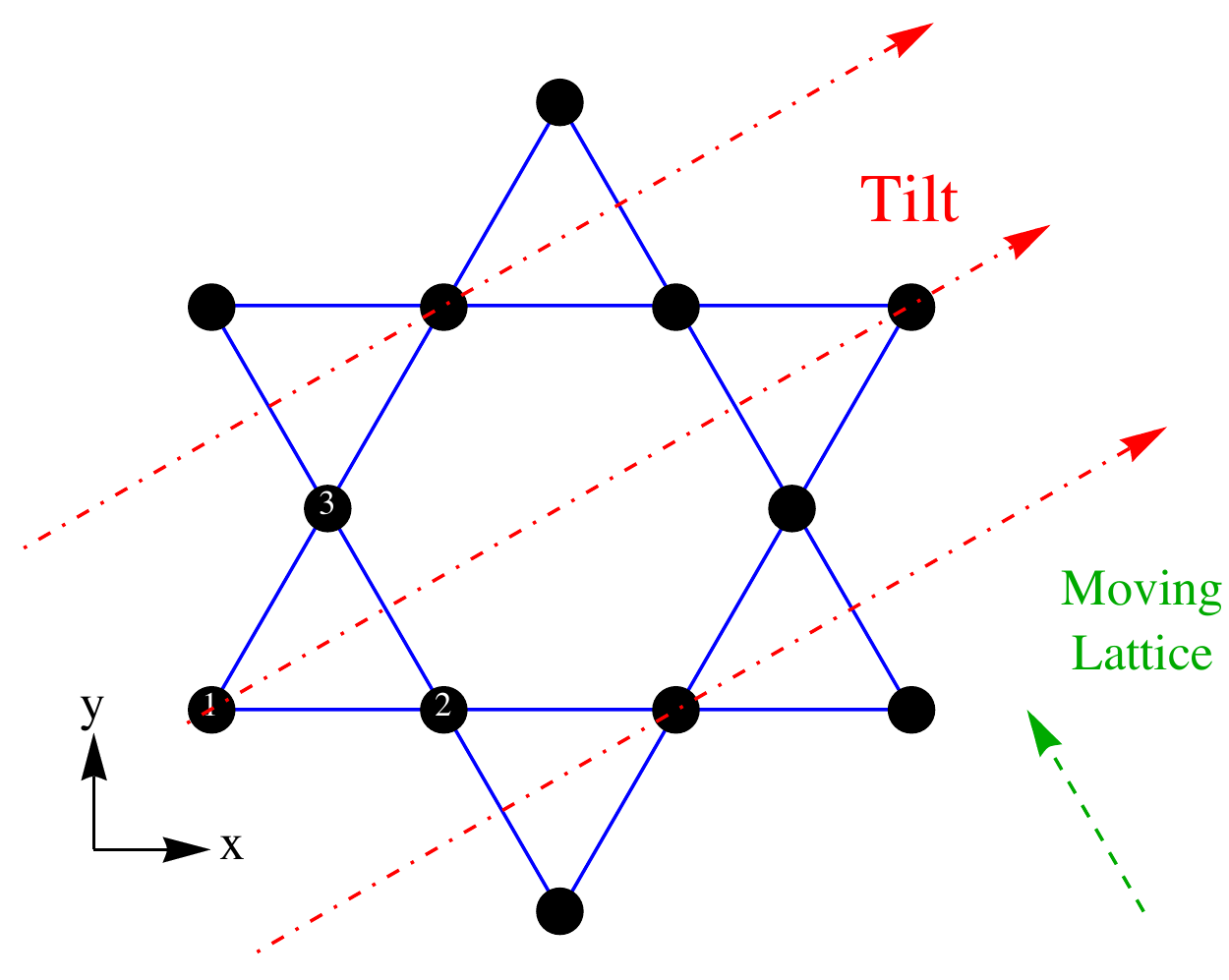}
	\end{center}
	\caption{Schematic of a kagome optical lattice with two additional fields applied to create a staggered  effective flux (Fig.~1b).  The arrows denote a uniform potential gradient (tilt) from a gravitational field, magnetic field, or another method.  The green arrows denote the direction of the moving lattice created by additional Raman beams.   }\label{fig_configuration}
\end{figure}
 
 Figure~\ref{fig_configuration} shows a schematic of the kagome lattice containing two species of fermion, e.g., $^{40}$~K, with both external fields applied, the tilt, and the moving lattice.  The kinetic energy of the atoms under the applied fields becomes a function of time $\tau$:
\begin{equation}
\tilde{H}_0^{\text{St}}=-t\sum_{\left<ij\right>\sigma}c_{i\sigma}^\dag c_{j\sigma}^{\vphantom{\dag}}
+\sum_{i\sigma}\left[\vec\Delta\cdot\bm R_i
+V(\bm R_i,\tau)\right] c_{i\sigma}^\dag c_{i\sigma}^{\vphantom{\dag}},
\label{eq_H_tilt_moving}
\end{equation}
where the annihilation and creation operators refer to undressed fermions (as in Eq.~2 but prior to the applied fields) in Wannier states localized at sites $i$ and $j$.  The second term results from the tilt field $\vec\Delta=\Delta\left[\hat x+(\sqrt 3/3)\hat y\right]/2$ such that $\vec\Delta\cdot\bm R_1=0$ and $\vec\Delta\cdot\bm R_2=\vec\Delta\cdot\bm R_3=\Delta/2$, where we assume the positions of the three sites in a unit cell are $\bm R_1=\bm 0$, $\bm R_2=\hat x$ and $\bm R_3=(1/2)\hat x+(\sqrt3/2)\hat y$.  The last term is due to a moving lattice created by additional lasers added to the lasers defining the kagome potentials: $V(\bm r,\tau)=\Omega\sin(\bm P\cdot \bm r-\tau \Delta)$, where $\bm P=-(\pi/2)\hat x+(\sqrt3\pi/2)\hat y$ is the momentum of the moving lattice, such that $\bm P\cdot\vec\Delta=0$, and $\bm P\cdot\bm R_3=\bm P\cdot\bm R_1+\pi/2=\bm P\cdot\bm R_2+\pi$.  Here we have chosen an oscillation frequency that helps maximize flux and equalizes the magnitude of the hopping along all bonds.

We derive a steady state effective model for the fermions under the applied fields.  By computing the Wannier functions in the presence of the tilt we find a Wannier-Stark effect which allows the moving lattice to generate a complex hopping (see Appendix~\ref{app:wannier}).  
Appendix~\ref{app:staggered} shows that the flux through the kagome lattice is staggered (Fig.~\ref{fig_schematics_kagome_flux}b).  Specifically, we find that a tilt and moving lattice applied to fermions in a kagome optical lattice results in Eq.~\ref{eq_H_0_end} with complex hopping and staggered flux to yield $\Phi_{\Delta}=-\Phi_{\nabla}=\Phi_0/4$.  We have checked that varying the angle and other parameters does not lead to a uniform flux, though other irregular flux patterns are possible.  We conclude that a method already realized in the laboratory (introducing flux in optical lattices using a tilt and a moving lattice) always leads to staggered flux patterns in kagome optical lattices. 
 
 We now turn to interaction effects in the staggered flux case.  We again take the Hubbard limit of a deep optical lattice with one particle per site. Arguments discussed in Sec.~\ref{sec:hubb} lead to a spin model of the form:
  \begin{eqnarray}
 &H^{\text{St}}&\approx J_{\text{H}}\sum_{\left<ij\right>} \bm S_i\cdot \bm S_j \nonumber \\
&+&J_{\text{C}}(\Phi_0/4) \biggl [
\sum_{ijk\in\bigtriangleup} \bm S_i\cdot(\bm S_j\times\bm S_k) 
-\sum_{ijk\in\bigtriangledown}\bm S_i\cdot(\bm S_j\times\bm S_k)
\biggr] \nonumber \\
&+& \mathcal{O}(t^4/U^3)
\label{eq_spin_model_staggered}
\end{eqnarray}
 where we assumed a staggered flux, $\Phi_{\Delta}=-\Phi_{\nabla}=\Phi_0/4$, that leads to a three-spin interaction term that alternates from triangle to triangle and $J_{\text{C}}(\Phi_0/4)=24t^3/U^2$.

 We now speculate on the role of strong interactions in the staggered flux case.  The staggered flux ground state of Eq.~\ref{eq_spin_model_staggered} is argued \cite{Bauer2018a} to be a gapless spin liquid where the zero-energy excitations fall along three nodal lines that all cross zero in momentum space.  The gapless nodal lines are protected by symmetry but finite size effects may open a gap.  We used numerical exact diagonalization on Eq.~\ref{eq_spin_model_staggered} with up to 18 spins with periodic boundary conditions to study the spectrum in all spin sectors.  We find small gaps ($\gtrsim 0.05  J_{\text{C}}$) at expected gapless points.  We conclude that large system sizes are needed to see the degeneracy because the gapless spectrum allows strong finite size effects.  Numerical work on kagome ladders with as many as 200 spins show a gap \cite{Bauer2018a} that decreases linearly with system size from $\sim 0.012 J_{\text{C}}$ for 50 spins to below $0.002 J_{\text{C}}$ for 200 spins, thus establishing a gapless phase for large system sizes.  This work also shows that the gapless phase is stable for $J_{\text{C}}/J_{\text{H}}\gtrsim 0.8$.   This range corresponds to $t/U \gtrsim 2/3$ and $ \vert \Phi_{\Delta,\nabla}\vert=\Phi_0/4$ in terms of Hubbard parameters, indicating that the gapless spin liquid phase is indeed reachable in a perturbative limit where $t/U$ is still less than one.   Further work would be needed to study the gapless phase for lower values of $t/U$, where the perturbative limit is more precise. 

\section{Conclusion}\label{sec:discussion}

Fermions in a kagome optical lattice in the Heisenberg limit can be driven into QSLs by applying fluxes that lead to chiral three-spin terms. If the final state of the combined lattice/flux system is to approximate a thermal state, we must assume that the initial state is at low enough entropies \cite{McKay2011} to lead to an approximation to the QSLs discussed here.  Recent work estimates that entropies per particle below $\sim 0.8 k_{\text{B}}$ are needed to reach the Laughlin regime of bosons \cite{Raum2017}, which is closely related to the CSL.  The entropy to reach the bosonic Laughlin state is within reach of atomic gas microscopes \cite{Bakr2009,Sherson2010,Bakr2010,Endres2011,Weitenberg2011a,Islam2015,Hild2014,Preiss2015,Cheuk2015,Parsons2015,Haller2015,Miranda2015a,Yamamoto2016,Parsons2016,Boll2016,Cheuk2016,Choi2016,Drewes2017,Mazurenko2017} which have already realized the Heisenberg (antiferromagnetic) limit in a square optical lattice \cite{Mazurenko2017} with entropies per particle below $\log(2) k_{\text{B}} \approx 0.7 k_{\text{B}} $.  To make such an estimate for the CSL, a detailed study of the high energy statistics of the CSL would be needed to extract an entropy-temperature relationship.  Such a study is beyond the scope of the present work.

The gapless spin liquid, by contrast, hosts a large number of (nearly) zero-energy  states and may therefore offer favorable entropy requirements.  The required entropy (which scales as the logarithm of the number of ground states) is not as low as the CSL.  A single spin excitation along one of the degenerate nodal lines hosts an entropy per particle $\sim\log(N)/N k_{\text{B}}$, for $N$ spins.  High occupancy of degenerate nodal lines implies that entropy can be large in finite sized systems.  From an entropy perspective, gapless spin liquids therefore appear to be simpler to realize because the low energy manifold can be accessed at higher entropies in finite sized systems.  

QSL ground states discussed here are more difficult to observe than conventionally ordered spin states (e.g., antiferromagnetic or ferromagnetic states \cite{Hofstetter2002,Mazurenko2017,Koutentakis2019}) because the QSL ground states are uniform and otherwise featureless.  The most obvious route to observe the CSL is the gap, manifest in the energy cost to change the spin imbalance.  The absence of a net magnetization but an observable gap would offer strong evidence for a CSL.  Additionally,  the CSL has chiral edge modes which could be observable using the spin analogue of recently realized quantized circular dichroism \cite{Tran2017,Asteria2019}. 
Spin liquids also distinguish themselves in their excitations. CSLs have anyon excitations which can lead to non-trivial power law behavior \cite{Morampudi2017,Yao2018} in the dynamical structure factor and can be observed with Bragg scattering \cite{Weidemuller1995,Ernst2010,Mottl2012,Hart2015}.  More local probes can be used to directly observe anyons in optical lattices captured by spin models \cite{Zhang2007b}. The gapless spin liquid phase can be revealed in measures of the dynamical structure factor as excitations populate degenerate nodal lines, revealing the gapless spinon surfaces. In this paper we have constructed a route to such spin liquids in ultra-cold atom systems with short-ranged interactions to foster their identification in the laboratory.   

\begin{acknowledgments}
V.W.S. acknowledges support from AFOSR (FA9550-18-1-0505).  V.W.S. and S.T acknowledge support from ARO (W911NF-16-1-0182).   We thank B. Bauer for helpful conversations. 
\end{acknowledgments}

\appendix

 
\section{Wannier-Stark States and the Effective Hamiltonian}
\label{app:wannier}

\begin{figure}[t]
	\begin{center}
		\includegraphics[width=0.48\textwidth]{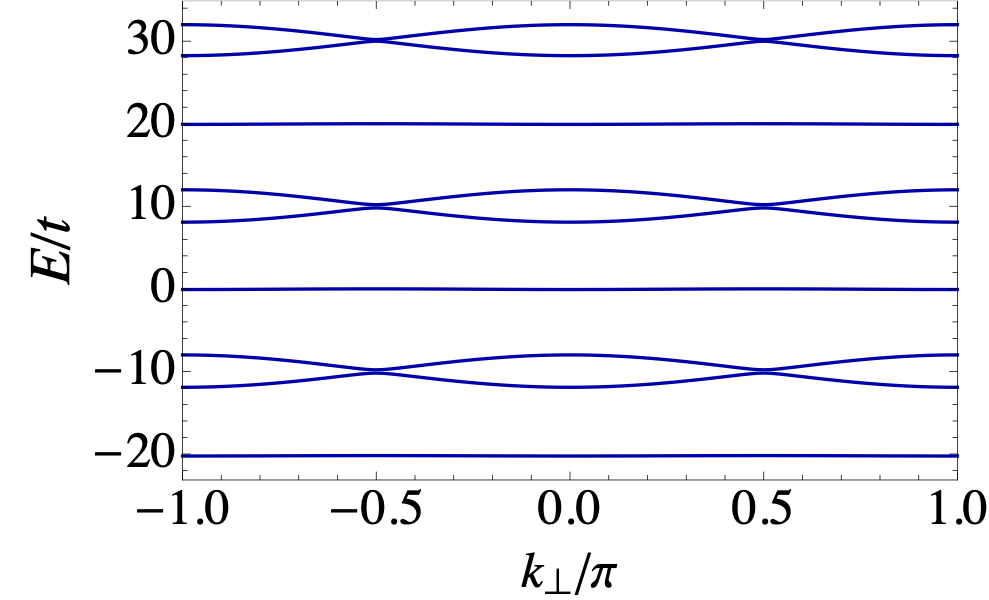}
	\end{center}
	\caption{Eigenvalues of the static part of Eq.~\ref{eq_H_tilt_moving} ($\Omega=0$) plotted against lattice momentum perpendicular to the tilt for $\Delta=20t$.  The kagome lattice is infinite along the direction of the moving lattice but extends three unit cells along the direction of the tilt, Fig.~\ref{fig_configuration}.  The dispersive (flat) bands define de-localized (localized) states used to vary hopping around triangles.
	}\label{fig_wannier_stark}
\end{figure}

In this section we show that Eq.~\ref{eq_H_tilt_moving} leads to Eq.~\ref{eq_H_0_end} with complex hopping and effective flux.  To find a parameter regime yielding an effective flux from a combination of a tilt and a moving lattice, we first study the impact of the first two terms in Eq.~\ref{eq_H_tilt_moving} on the basis of Wannier functions.  We numerically solve for the eigen-modes for a system which is finite along the direction of tilt while infinite along the orthogonal axis. The momentum along the orthogonal axis, $k_\perp$, is a good quantum number.  In the limit of strong tilt, we find two types of states plotted in  Fig.~\ref{fig_wannier_stark}. States localized near site $\bm R_1$ are dispersionless since their hopping to sites $\bm R_2$ and $\bm R_3$ (See Fig.~\ref{fig_configuration}) are suppressed due to the energy difference, and there is no hopping possible along the direction perpendicular to the tilt. These states appear as flat bands in Fig.~\ref{fig_wannier_stark}. (This suppression of hopping is key to allowing the moving lattice to generate a complex hopping.)  States localized near sites $\bm R_2$ and $\bm R_3$ can hop freely along the direction perpendicular to the tilt and therefore form the dispersive bands in Fig.~\ref{fig_wannier_stark} with bandwidth $4t$.  Wannier-Stark states are then constructed from the Fourier transform of the Bloch states, where the phases are chosen to yield states maximally localized on a lattice site.
We denote the Wannier-Stark states localized near $\bm R_i$ by $\left|i\right>$.

In the basis of Wannier-Stark states, the Hamiltonian in the presence of both the tilt and moving lattice becomes:
$
h_0^{\text{WS}}= -t\sum_{\left<ij\right>}^{'}\left|i\right>\left<j\right|+\sum_i\left(\vec\Delta\cdot\bm R_i\right)\left|i\right>\left<i\right|+\Omega\sum_{ij}\left|i\right>\left<i\right|\sin(\bm P\cdot\bm r- \tau \Delta)\left|j\right>\left<j\right|,
$
where the prime on the sum indicates a sum only over bonds such that $\vec\Delta\cdot(\bm{R} _i-\bm{R}_j)=0$.  To remove the time dependence we pass to the rotating basis, defined by the unitary time evolution operator:
$
U=\exp\left\{i\sum_i\left[-\left(\vec\Delta\cdot\bm R_i\right)\tau-F_{\bm P,\bm r}\cos(\bm P\cdot\bm R_i-\tau\Delta )\right]\left|i\right>\left<i\right|\right\}, 
$
where 
$F_{\bm P,\bm r}\equiv(\Omega/\Delta)\left<0\right|\cos\bm P\cdot\bm r\left|0\right>$.  Using $U$ we can now remove the time dependence in $h_0^{\text{WS}}$ using:
 $U^\dag h_0^{\text{WS}} U-iU^\dag(\partial U/\partial t)$.  The resulting model is time independent but now describes dressed fermions with complex hopping, i.e, we retrieve Eq.~\ref{eq_H_0_end}.  Direct numerical simulation of the Wannier functions and computation of the resulting imaginary part of the hopping shows that staggered flux with tunable strength is possible.  An analytic argument for the staggered flux pattern can be derived in the weak $\Omega$ limit.\\
 
 \section{Staggered Flux in the Weak $\Omega$ Limit }
\label{app:staggered}

In this section we show that the flux derived from Eq.~\ref{eq_H_tilt_moving} is staggered.  We have computed this numerically in a tight-binding construction of the complex hoppings in Eq.~\ref{eq_H_0_end}. We can work in the weak $\Omega$ limit to allow analytic expressions demonstrating the mechanism behind the staggered flux. If we let indices, 1,2, and 3 refer to the sites in upward-pointing triangle in Fig.~\ref{fig_configuration} we find (for weak $\Omega$) a complex hopping:
$
t_{1n}\approx\Omega\left<1\right|e^{-i\bm P\cdot\bm r}\left|n
\right>/(2i)
$
for $n=2,3$, 
and real hopping along the remaining bond in the triangle:
$
t_{23}
\approx t {\cal J}_0\left[(2\Omega/\Delta)\sin(\bm P\cdot\bm R_{23}/2)\right].
$
 We have checked using maximally localized Wannier functions that we can maximize the flux through the plaquettes and adjust $\Omega$ to set  $t=\vert t_{12}\vert=\vert t_{23}\vert=\vert t_{31}\vert$, leading to $\text{Im}[ t_{12} t_{23} t_{31} ]= t^3 \sin(2\pi\Phi_{\Delta}/\Phi_0)\approx t^3$.  This shows that we can use the moving lattice to induce an effective flux in the kagome lattice.
 
 The flux for the downward triangles is different. One can show that $t_{1n}\sim \exp[i \bm P \cdot (\bm R_1+\bm R_n) ]$ for weak $\Omega$ while $t_{23}$ is real and the same for all triangles.  This implies that the sign of the flux is the  \emph{opposite} for downward pointing triangles in comparison to upward pointing triangles in Fig.~\ref{fig_configuration}, i.e.,  
 $t^3 \sin(2\pi\Phi_{\nabla}/\Phi_0)\approx -t^3$.  The change in the sign of the flux arises from the change in sign of the moving lattice potential set by $\bm P$.  This behavior contrasts with the uniform flux realized using the same technique but in square optical lattices.

\bibliography{references}

\end{document}